# Quantum microwave photonics

## Authors

Ye Yang,[1,3,5] Yaqing Jin,[2,4] Xiao Xiang,[2,4] Wei Li,[1,3,5] Tao Liu,[2,4] Shougang Zhang[2,4] Ruifang Dong[2,4,*] and Ming Li[1,3,5#]

## Affiliations

[1]State Key Laboratory on Integrated Optoelectronics, Institute of Semiconductors, Chinese Academy of Sciences, Beijing, 100083, China

[2]Key Laboratory of Time and Frequency Primary Standards, National Time Service Center, Chinese Academy of Sciences, Xi'an 710600, China

[3]School of Electronic, Electrical and Communication Engineering, University of Chinese Academy of Sciences, Beijing 100049, China

[4]School of Astronomy and Space Science, University of Chinese Academy of Sciences, Beijing 100049, China

[5]Center of Materials Science and Optoelectronics Engineering, University of Chinese Academy of Sciences, Beijing 100190, China

Corresponding author: [*]dongruifang@ntsc.ac.cn; [#]ml@semi.ac.cn

**By harnessing quantum superposition and entanglement, remarkable progress has sprouted over the past three decades from different areas of research in communication [1,2,3,4], computation [5,6,7] and simulation [8,9]. To further improve the processing ability of microwave photonics, here, we have demonstrated a quantum microwave photonic processing system using a low jitter superconducting nanowire single photon detector (SNSPD) and a time-correlated single-photon counting (TCSPC) module. This method uniquely combines extreme optical sensitivity, down to a single-photon level (below −100 dBm), and wide processing bandwidth, twice higher than the transmission bandwidth of the cable. Moreover, benefitted from the trigger, the system can selectively process the desired RF signal and attenuates the other intense noise and undesired RF components even the power is 15dB greater than the desired signal power. Using this method we show microwave phase shifting and frequency filtering for the desired RF signal on the single-photon level. Besides its applications in space and underwater communications [10,11] and testing and qualification of pre-packaged photonic modulators and detectors [12]. This RF signal processing capability at the single-photon level can lead to significant development in the high-speed quantum processing method.**

Microwave photonics (MWP) typically using classic optical methods and devices to generate, transport, and process radio-frequency (RF) signals [13,14,15] is a recently developed area of research which has greatly improved the microwave processing system in frequency, bandwidth, dynamic range and anti-interference. However, increasing bandwidth demands a higher electrical transmission bandwidth and a larger optical power at the photodetector because of the gain-bandwidth product of the detector. Considering that the photonic quantum technologies are adept at manipulating and processing the photon signal with extremely low optical power, further acquiring improvement for the MWP processing is expected by introducing the photonic quantum devices.

In this article, we propose a quantum MWP system combining the single-photon technology and MWP. In this system, though the continuous high-intensity optical signal is transformed to the discretely-distributed single-photon signal, the original microwave photonic processing method [16] is unaffected. For the backend detection, the low-jitter SPD transforms the required high-speed RF response to the received optical signal into a slow response to the detected single photons. Combined with the TCSPC method [17, 18], the desired high-speed RF signal can be recovered from the single-photon signal. Therefore, the quantum MWP link can be conveniently and effectively used in processing high frequency RF signal in the backend transmission system without considering the transmission bandwidth. Compared with the other high-bandwidth MWP methods based on optical mixing[19], amplification[20], sampling[21], dispersive[22, 23], and nonlinear techniques[24], the simplicity and calibration-free nature of the method, as well as its capability of selectively processing the RF signal, such as phase shifting and frequency filtering shown here make it a more viable solution to process the high-speed waveform, and lays the foundation for further processing RF signal based on quantum methods.

Figure 1 illustrates the operation principle. The sliced carrier was amplitude modulated using the electro-optic modulator driven by high-speed RF signals. The optical waveform then was attenuated down to a single-photon flux level. After dispersed the modulated single-photon signal was detected by a superconducting nanowire single photon detector (SNSPD) [25]. The SNSPD's response is transmitted to the TCSPC as the "stop" pulses through the cable. The photon capturing speed of the SNSPD is much slower than the high-speed RF signal. Meanwhile, a slow periodic frequency-locked electrical trigger was sent to the TCSPC as the "start" signal. Using the TCSPC sampling acquisition method, the SNSPD's dead-time limits for capturing high-speed signal can be overcome [26, 27]. Finally, the high-speed RF signal is recovered from the recorded periodic-varying counts as a function of the start-stop time difference. Due to the low jitter of the SNSPD and TCSPC, our system achieves the ability to recover the high-speed RF signal that is far beyond the transmission bandwidth of the electronic backend instrument and with a single-photon sensitivity [26, 27, 28]. By using the purposed quantum MWP system, the high-performance RF phase shifting, frequency filtering and the signal selectivity in noisy radio environments has been demonstrated.

Figure 2 presents the high-speed single-photon phase shifting performance. We use two DCM as the dispersion media. The frequency of the involved RF signal $f_{\rm rf}$ is chosen as 5 GHz and 8 GHz respectively, which are both above the bandwidth of the TCSPC system and the cable (around 3 GHz). Figure 2a presents the recovered photon count waveforms from the TCSPC. The dashed line is the fitting curve of corresponding measured counts, ignoring the noise-like deviation on time bins. In case A and B, the DCM1 is used to provide the dispersion while the RF frequency $f_{\rm rf}$ is chosen as 5 GHz and 8 GHz respectively. The chosen RF frequencies $f_{\rm rf}$ in case C and D are as same as case A and B but using DCM2 to apply the dispersion. As shown in Fig. 2a, the system successfully changes the initial phase of a high-speed single-photon output by adjusting the carrier's center wavelength, which is similar to the performance of the traditional MWP phase shifter. Though the recovered signal for the higher frequency at 8 GHz has a degraded contrast, we should point out that improvement can be achieved by using the single sideband modulator and the SNSPD with lower jitter. Calculated from the initial phases, the relative delay as a function of the carrier wavelength are plotted in dots in Fig. 2b. The slopes of the relative delay (from case A to case D) are $-325.11(\pm 6.61)$ps/nm, $-333.69(\pm 15.22)$ps/nm, $-494.42(\pm 18.36)$ps/nm and $-503.43(\pm 9.72)$ps/nm respectively, which are highly consistent with the claimed dispersion values of the DCM1 and DCM2. The result indicates that the signal conversion in our system performs exactly the same phase-shifting function as the traditional method and further demonstrates the high performance for recovering the high-frequency RF signal at the single-photon level.

Consider more than one carrier with a different center wavelength passes through the system, the recovered output signal on different carriers interfere with each other, and the frequency filtering function on the RF signal can be achieved. In the frequency filtering experiment, the DCM2 is used. We use two carriers modulated by RF signal passing through our system. The two carriers' center wavelengths in Fig. 3a are 1550 nm and 1552 nm. When $f_{rf} = 1.00$GHz, the amplitude of the measured waveform with the DCM is close to the case without the DCM. However, when $f_{rf} \approx 1.5$GHz, the amplitude of the measured waveform with the DCM is significantly lower than the result without the DCM, which indicates an interference cancellation between the outputs based on two carriers. The contrast ratio between the cases with and without the DCM is plotted in Fig. 3b when the RF frequency is ranged from 100 MHz to 3 GHz, and a ratio of more than 15 dB is observed. Similarly with the traditional MWP link, our system is continuously tunable by adjusting the number and the wavelengths of the photon carriers. By setting the carriers' wavelengths to 1550 nm and 1553 nm, another response is presented. By comparing the results, the FSR tenability is indicated by adjusting the carriers' wavelength gap. Further adjustment of the filtering shape can be achieved by increasing the number of the photon carrier. To test our process-ability for high-frequency RF signal, Fig. 3c records the contrast ratio for the signal from 5 GHz to 8 GHz. When the RF frequency is approaching 8 GHz, the ratio becomes lower than the theoretical curve, which can also be attributed to the jitter of the SNSPD.

Note that the phase-stabilized trigger is important to selectively recover the specific photon waveform. To verify the selective process, we provide another RF signal without phase locking, which cannot be recovered like the former results. Therefore the undesired RF component degrades the input signal to noise ratio (SNR). To test the system's selectively processing ability, we record the recovered signal with different signal power and dispersion when the power of the undesired input is at 9 dBm. With the DCM1 as the dispersion medium, the measured waveforms are plotted in Fig. 4a (form case A to case D). We define the "noise", which is plotted in yellow, as the deviation of the counts (blue) to the fittings (red) after subtracting the mean photon counts. As can be seen in Fig. 4a, the amplitude of the acquired waveform becomes smaller when the desired signal's power decreases, while the noise signal remains the same.

Figure 4b and 4c plot the waveform amplitude and corresponding "SNR" of the desired signal's output when the input signal power increase from -12 dBm to 9 dBm while the power of the undesired input is kept as 9 dBm. The influence of different dispersion ranging from 0 to 495 ps/nm is investigated and shown as well. As Fig. 4b points out, the amplitude of the output signal increases with the signal power, but higher GVD will decrease the amplitude. The output "SNR" (Fig. 4c) indicates that the specific signal with "SNR" larger than 0 dBm can be selected even the $P_{ud}$ is at least 17 dB larger than the $P_{rf}$. We should point out that a better selective performance can be achieved by increasing the accumulation time for the recovering process. The results indicate a good selectivity and resistance when the signal is mixed with other undesired components. Compared with the results in different $f_{ud}$, the characteristics of the waveform, such as the pattern, initial phase, and the "SNR", are highly comparable, which indicates that the frequency of the undesired input does not affect the signal processing and capturing. Such performance can be attributed to the stable in-phase relationship between the desired signal and the trigger. Therefore, this system has a strong immunity to undesired intense RF white noise as long as it is out-of-phase with the trigger.

In summary, we have demonstrated a quantum microwave photonic processing system by combining the MWP link with the single-photon technology. With the help of the low jitter SNSPD and TCSPC, the processing system has both extreme sensitivities, down to a single-photon detection event, and wide processing bandwidth. Through the conversion between the high-speed optical waveform and the single-photon flux signal, our system, overcoming the electronic bandwidth after the SNSPD, can process the wide-bandwidth RF signal, whose frequency is significantly higher than the transmission bandwidth. Moreover, relying on the trigger clock, the system effectively attenuates the intense noise and selectively processes the desired RF input. Besides the potential application in high loss system, our system can be highly potential to process high-speed RF signals with the quantum processing method. For example, such quantum microwave photonic can further combine with the entangle feature to illustrate the nonlocal or secure quantum microwave processing.

**Data availability:**

The data that support the findings of this study are available from the corresponding author upon reasonable request.

**Acknowledgements**

This work was supported by National Key Research and Development Program of China under 2018YFB2201902, 2018YFB2201901 and 2018YFB2201903. This work was also partly supported by the National Natural Science Foundation of China under 61925505, 61535012 and 61705217

**Disclosures**

The authors declare no conflicts of interest.

**Methods**

**Experimental details:**

The broadband ASE source operating at C-band is spectrally sliced by the programmable pulse shaper (Finisar, WaveShaper 4000S) to provides optical sliced carriers. The carrier was directly amplitude modulated using the electro-optic modulator (EOSPACE) driven by high-speed RF signals from the signal generator (KEYSIGHT, E8257D). The dispersion is provides by the dispersion compensation module (DCM, Proximion DCM-CB). In our experiment, two DCMs (DCM1 and DCM2) are used. The GVD is -330 ps/nm and -495 ps/nm. The utilized SNSPD in our experiment is not a commercial product. It was manufactured by Prof. Lixing You's group from the Shanghai Institute of Microsystem and Information Technology, Chinese Academy of Sciences, China. The details about the SNSPD can be found in ref[29]. The TCSPC (PicoQuant, PicoHarp 300) is used to record the start-stop time difference. The time-bin resolution of the TCPSC is set as 8 ps and the measurement time for every measured photon waveforms is 60 seconds. The electric cable is the normal cable with BNC connectors whose bandwidth is about 3 GHz. The phase-stabilized trigger is time base output of the signal generator (KEYSIGHT, E8257D) whose frequency is 10 MHz (It should be larger than the photon count rate by the SNSPD).

Considering the signal generator that we used in the experiments cannot provide noise with adjustable power and frequency. And the noise bandwidth (about 40MHz) is much smaller than the frequency of the input signal. In the experiment of selective processing (Fig. 4), the undesired RF signal without phase locking is from another independent signal generator (SRS, SG382)

**Reference:**

1. Bennett CH, Brassard G. Quantum cryptography: Public key distribution and coin tossing. *Proceedings of the IEEE International Conference on Computers, Systems and Signal Processing*. IEEE New York; 1984. pp. 175–179.

2. Bennett CH, Brassard G, Crépeau C, Jozsa R, Peres A, Wootters WK. Teleporting an unknown quantum state via dual classical and Einstein-Podolsky-Rosen channels. *Phys Rev Lett* 1993, **70**(13): 1895.

3. Gisin N, Thew R. Quantum communication. *Nat Photonics* 2007, **1**(3): 165-171.

4. Lo H, Curty M, Tamaki K. Secure quantum key distribution. *Nat Photonics* 2014, **8**(8): 595-604.

5. DiVincenzo DP. Quantum computation. *Science* 1995, **270**(5234): 255-261.

6. Shor PW. Polynomial-Time Algorithms for Prime Factorization and Discrete Logarithms on a Quantum Computer. *SIAM J Comput* 1997, **26**(5): 1484-1509.

7. Albash T, Lidar DA. Adiabatic quantum computation. *Rev Mod Phys* 2018, **90**(1): 015002.

8. Lloyd S. Universal Quantum Simulators. *Science* 1996, **273**(5278): 1073-1078.

9.  Aspuru-Guzik A, Walther P. Photonic quantum simulators. *Nat Phys* 2012, **8**(4)**:** 285-291.

10. Hemmati H, Biswas A, Djordjevic IB. Deep-Space Optical Communications: Future Perspectives and Applications. *Proc IEEE* 2011, **99**(11)**:** 2020-2039.

11. Hanson F, Radic S. High bandwidth underwater optical communication. *Appl Opt* 2008, **47**(2)**:** 277-283.

12. De Coster J, De Heyn P, Pantouvaki M, Snyder B, Chen H, Marinissen EJ*, et al.* Test-station for flexible semi-automatic wafer-level silicon photonics testing.  european test symposium; 2016; 2016. p. 1-6.

13. Seeds AJ, Williams KJ. Microwave Photonics. *J Lightwave Technol* 2006.

14. Capmany J, Novak D. Microwave photonics combines two worlds. *Nat Photonics* 2007, **1**(6)**:** 319-330.

15. Yao J. Microwave Photonics. *J Lightwave Technol* 2009.

16. Capmany J, Ortega B, Pastor D. A tutorial on microwave photonic filters. *J Lightwave Technol* 2006, **24**(1)**:** 201-229.

17. McCarthy A, Krichel NJ, Gemmell NR, Ren X, Tanner MG, Dorenbos SN*, et al.* Kilometer-range, high resolution depth imaging via 1560 nm wavelength single-photon detection. *Opt Express* 2013, **21**(7)**:** 8904-8915.

18. Shcheslavskiy V, Morozov PV, Divochiy A, Vakhtomin YB, Smirnov K, Becker W. Ultrafast time measurements by time-correlated single photon counting coupled with superconducting single photon detector. *Rev Sci Instrum* 2016, **87**(5): 053117.

19. Donaldson WR, Marciante JR, Roides RG. An Optical Replicator for Single-Shot Measurements at 10 GHz With a Dynamic Range of 1800:1. *IEEE J Quantum Electron* 2010, **46**(2): 191-196.

20. Mahjoubfar A, Goda K, Betts G, Jalali B. Optically amplified detection for biomedical sensing and imaging. *J Opt Soc Am A* 2013, **30**(10): 2124-2132.

21. Takara H, Kawanishi S, Morioka T, Mori K, Saruwatari M. 100 Gbit/s optical waveform measurement with 0.6 ps resolution optical sampling using subpicosecond supercontinuum pulses. *Electron Lett*: Institution of Engineering and Technology; 1994. pp. 1152-1153.

22. Solli DR, Chou J, Jalali B. Amplified wavelength-time transformation for real-time spectroscopy. *Nat Photonics* 2008, **2**(1): 48-51.

23. Mahjoubfar A, Churkin DV, Barland S, Broderick NGR, Turitsyn SK, Jalali B. Time stretch and its applications. *Nat Photonics* 2017, **11**(6): 341-351.

24. Andrekson PA. Picosecond optical sampling using four-wave mixing in fibre. *Electron Lett* 1991, **27**(16):

1440-1441.

25. Wu J, You L, Chen S, Li H, He Y, Lv C, *et al.* Improving the timing jitter of a superconducting nanowire single-photon detection system. *Appl Opt* 2017, **56**(8): 2195-2200.

26. Wang X, Dane AE, Berggren KK, Shaw MD, Mookherjea S, Korzh B, *et al.* Oscilloscopic Capture of Greater-Than-100 GHz, Ultra-Low Power Optical Waveforms Enabled by Integrated Electrooptic Devices. *J Lightwave Technol* 2020, **38**(1): 166-173.

27. Wang X, Korzh B, Weigel PO, Nemchick D, Drouin BJ, Fung A, *et al.* Oscilloscopic Capture of 100 GHz Modulated Optical Waveforms at Femtowatt Power Levels.  optical fiber communication conference; 2019; 2019.

28. Fedder H, Oesterwind S, Wick M, Olbrich F, Michler P, Veigel T, *et al.* Characterization of Electro-Optical Devices with Low Jitter Single Photon Detectors – Towards an Optical Sampling Oscilloscope Beyond 100 GHz.  european conference on optical communication; 2018; 2018.

29. You L, Yang X, He Y, Zhang W, Liu D, Zhang W, *et al.* Jitter analysis of a superconducting nanowire single photon detector. *AIP Adv* 2013, **3**(7): 072135-072135.

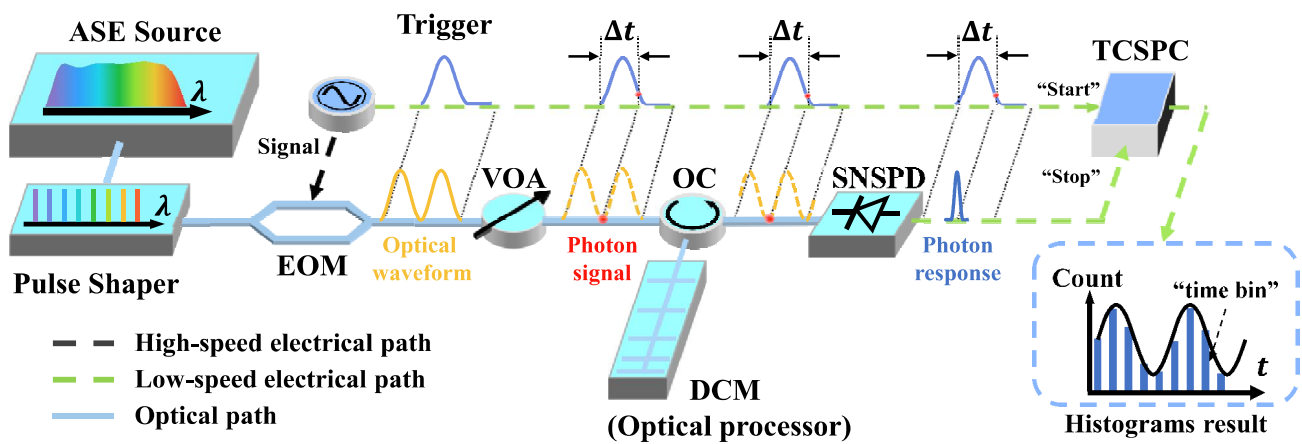

**Fig. 1 | Experimental setup of the quantum photonic microwave processing system.** A spectrally sliced optical carrier, modulated by high-speed RF signal, is transformed to single-photon flux via the variable attenuator (VA). The single-photon signal is firstly dispersed by the dispersion compensation module (DCM) and then detected by a superconducting nanowire single photon detector (SNSPD). With the SNSPD's response being the input of the TCSPC and the 10 MHz signal from the signal generator as its clock trigger, the periodic start-stop time difference is recorded to recover the high-speed RF signal.

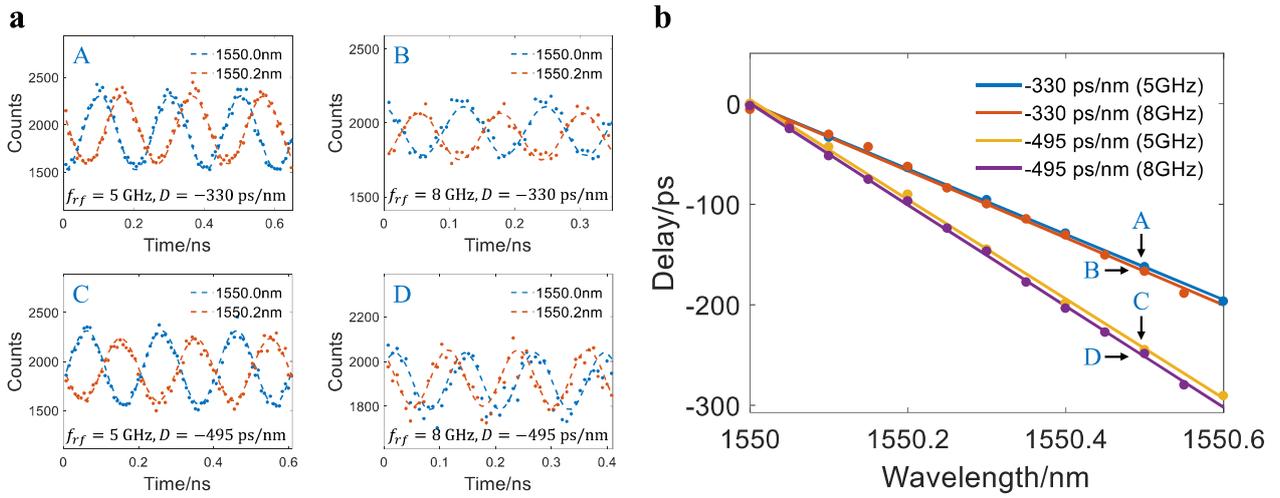

**Fig. 2 | The measurement results of the phase shifting. a,** the recovered signal at the TCSPC with different center wavelength, frequency and dispersion (case A,5 GHz@-330 ps/nm; case B, 8 GHz@-330 ps/nm; case C,5 GHz@-495 ps/nm; case D,8 GHz@-495 ps/nm). **b,** The fitting curve of the relative delay and the passband center wavelength which different RF frequency and GVD is applied.

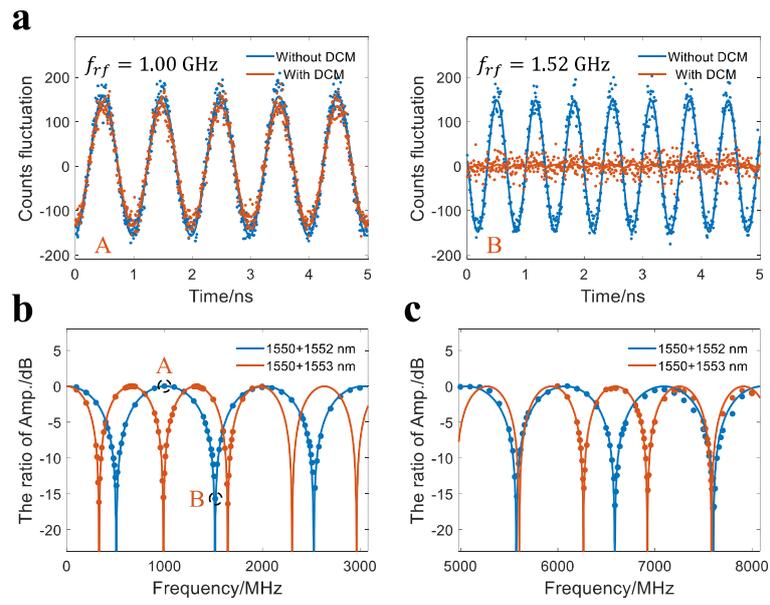

**Fig. 3 | The measurement results of the frequency filtering. a,** The recovered waveforms from the TCSPC with and without the DCM at different input frequency (case A, 1GHz; case B, and 1.5GHz). The frequency responses at different the wavelength gap between the photons and frequency range (**b**, 0.1-3 GHz; **c**, 5-8 GHz)

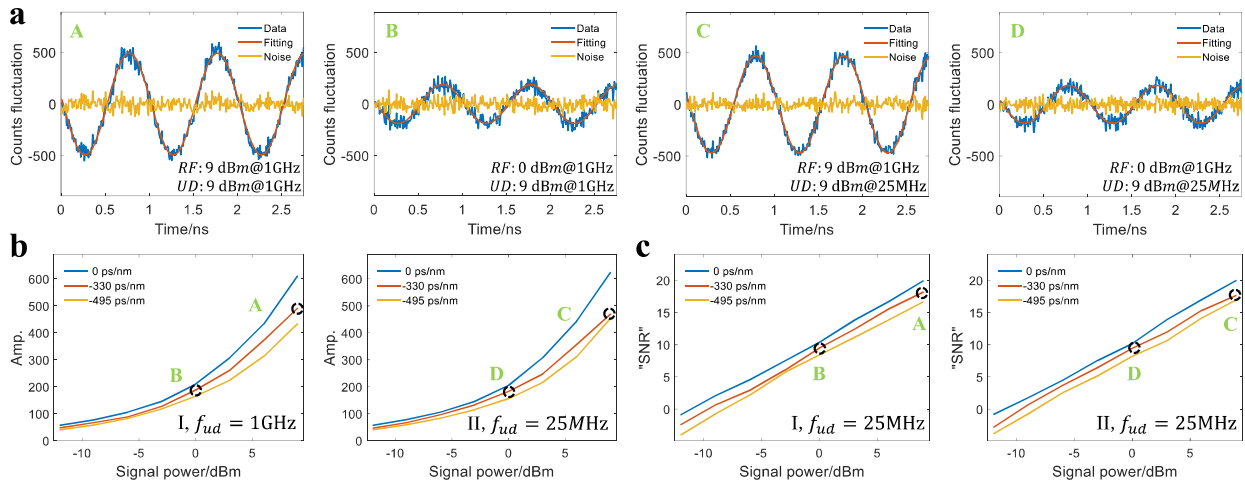

**Fig. 4 | The resistance to the undesired component. a**, the captured waveform of 1GHz signal with different power and frequency of undesired component (case A, $P_{rf} = 9dBm, f_{ud} = 1GHz$; case B, $P_{rf} = 0dBm, f_{ud} = 1GHz$; case C, $P_{rf} = 9dBm, f_{ud} = 25MHz$; case D, $P_{rf} = 0dBm, f_{ud} = 25MHz$). **b**, The fitted amplitude and **c**, the calculated "SNR" of the waveform for two different frequency of undesired component (case I, $f_{ud} = 1GHz$; case II, $f_{ud} = 25MHz$;) under different GVD conditions.